\ifx\mnmacrosloaded\undefined \input mn\fi

\pageoffset{-2.5pc}{0pc}
  
%\Autonumber  %  auto-number sections, subsections and subsubsections

% \pagerange, \pubyear and \volume are defined at the Journals office and
% not by an author.

\def\doublespace {\smallskipamount=6pt plus2pt minus2pt
                  \medskipamount=12pt plus4pt minus4pt
                  \bigskipamount=24pt plus8pt minus8pt
                  \normalbaselineskip=24pt plus8pt minus8pt
                  \normallineskip=2pt
                  \normallineskiplimit=0pt
                  \jot=6pt
                  {\def\smallskip {\vskip\smallskipamount}}
                  {\def\medskip   {\vskip\medskipamount}}
                  {\def\bigskip   {\vskip\bigskipamount}}
                  {\setbox\strutbox=\hbox{\vrule 
                    height17.0pt depth7.0pt width 0pt}}
                  \parskip 12.0pt
                  \normalbaselines}

\def\lax    {\ifmmode{_<\atop^{\sim}}\else{${_<\atop^{\sim}}$}\fi}
\def\gax    {\ifmmode{_>\atop^{\sim}}\else{${_>\atop^{\sim}}$}\fi}
\def\kms    {\ifmmode{{\rm ~km~s}^{-1}}\else{~km~s$^{-1}$}\fi}

\def\eg{{\it e.\thinspace g.~}}
\def\ie{{\it i.\thinspace e.~}}
\def\bk{\lower 6pt\hbox{${\buildrel k\over \sim}$}}
\def\bv{\lower 6pt\hbox{${\buildrel v\over \sim}$}}

\def\pdm {precision Doppler measurements~}

\def\rv {radial velocity~}
\def\b {binary~}
\doublespace

% \onecolumn        % enable one column mode
\letters          % for `letters' articles
\pagerange{L63--L66}    % `letters' articles should use \pagerange{Ln--Ln}
\pubyear{1997}
\volume{291}
% \microfiche{}     % for articles with microfiche
% \authorcomment{}  % author comment for footline

\begintopmatter

\title{Precision Measuring of Velocities
via the Relativistic Doppler effect}
\author{Leonid M. Ozernoy}
\affiliation{5C3, Computational Sciences  Institute and Department of Physics 
\& Astronomy, George Mason U., Fairfax, VA 22030-4444; also Laboratory for 
Astronomy and Solar Physics, NASA/Goddard Space Flight Center, Greenbelt, MD 
20771, USA; 
\hbox {e-mail: ozernoy@hubble.gmu.edu; ozernoy@stars.gsfc.nasa.gov}
}

\shortauthor{L.M. Ozernoy}
\shorttitle{Precise Velocity Measuring}

% \acceptedline is to be defined at the Journals office and not
% by an author.

\acceptedline{Accepted 1997 September 18. Received 1997 July 14;
  in original form 1995 November 3}

\abstract {Just as the ordinary Doppler effect serves as a tool to measure
radial velocities of celestial objects, so can the relativistic Doppler 
effect be implemented to measure a combination of radial and transverse 
velocities by using recent improvements in observing techniques. A key 
element that makes a further use of this combination feasible is the 
periodicity in changes of the orbital velocity direction for the source.
Two cases are considered: (i) a binary star; and (ii) a solitary star 
with the planetary companion. It is shown that, in case (i), several 
precision Doppler measurements employing the gas absorption cell technique
would determine both the total orbital velocity and the inclination 
angle of the binary orbit
disentangled from the peculiar velocity of the system. The necessary 
condition for that is the measured, at least with a modest precision, proper
motion and distance to the system.}


\maketitle  %  finish the two spanning material

\section{1. Introduction}

 Measuring transverse (tangential) velocities has always been one of the 
most important and, at the same time, the least advanced issues in astronomy.
Available methods include measurement of either annual parallaxes or proper 
motions, which is possible only for nearby objects. Furthermore, the method 
of proper motions requires knowledge of the distance to the object. This 
Letter shows that, just as the ordinary Doppler effect serves as a tool to 
measure radial velocities of celestial objects, the relativistic Doppler 
effect can be implemented to measure 
a combination of radial and 
transverse velocities  by using the rapidly improving technique of  
precision Doppler measurements. This combination, as I show here, makes it 
possible to derive the inclination angle of a binary star provided that 
both the distance to the binary and its proper motion are measurable.

\section{2. The Method}
\doublespace

The frequency of a spectral line measured at the Earth and corrected for the 
Earth's rotation is supposed to be associated with the geocentric coordinate
system. It is necessary to introduce at least two more reference frames: 
the frame of the source barycentre, $S'$, and the frame of the Solar 
system barycentre, $S''$. Let the motion of the source relative to $S'$ be
described by a (generally variable) velocity
$\vec \beta\equiv \vec v/c$,  $S'$ moves relative to $S''$ with a 
(constant in both  direction and time) velocity
$\vec b=\vec V/c$, and the Earth moves relative to $S''$ with
a velocity $\vec\beta_\oplus$.    
The relativistic Doppler equation that relates the observed  frequency 
of a spectral line, $\nu$, to the emitted frequency, $\nu_0$, is 
given by successive application of the  usual 
4-wavevector transformation (Landau \& Lifshitz 1951)
between all the reference frames involved:
$${\nu \over \nu_0} = {\nu \over \nu'} {\nu' \over \nu''} {\nu'' \over 
\nu_0} ={\sqrt {1-\beta^2} \over 1-\beta_r}{\sqrt {1-b^2} \over 1-b_r}
{1-\beta_\oplus \cos\theta \over
\sqrt{1-\beta_\oplus^2}}\equiv {\cal B}{\cal C}{\cal E},  \eqno{(1)}
$$ 
where  ${\cal B}$ absorbs the source's internal velocities $\beta$ and 
$\beta_r$, ${\cal C}$ 
absorbs the motion of the source barycentre with velocity $\vec b=
\vec V/c$, ${\cal E}$ absorbs the Earth's velocity $\vec \beta_\oplus$,
and $\theta $ is the angle in $S''$ between $\vec \beta_\oplus$ and  the 
direction to the source.

At first sight, the very structure of equation (1) -- an isolation of 
different types of velocity strictly within the corresponding 
multipliers, i.e. factorization -- makes it impossible to determine the
total and radial velocities separately. However, special circumstances, 
such as periodic changes 
of the direction of the source's velocity, improve the situation.
Below, I consider two particular cases: (i) the 
source as a binary star and (ii) the source as a star with the planetary
companion.

\subsection{2.1. Case (i): a binary}

Let us consider a \b in a circular orbit with inclination angle $i$.
In this case, the change of the \rv $\beta_r$  is periodic
while the total velocity $\beta=\mid\vec\beta\mid$ is constant. Then 
${\cal B}$ changes in a periodic fashion as well. Let $\beta_r$
reach its maximum, $\beta_{r,\rm max}$, at an instant $t_1$ so that 
${\cal B}$ has a maximum ${\cal B}_{\rm max}$ at that instant too. In half a 
period
(the instant $t_2$), $\beta_r=-\beta_{r,\rm max}$ so that ${\cal B}$ reaches 
its minimum ${\cal B}_{\rm min}$. Evidently, 
$${\cal B}_{\rm max}^{-1}+{\cal B}_{\rm min}^{-1}={2\over\sqrt {1-\beta^2}}, 
\eqno (2)  
$$
i.e. the dependence on $\beta_{r,\rm max}$ is cancelled out and the result
depends on $\beta^2$  only. Once  $\beta^2$ is found, $\beta_{r,\rm max}$ 
can be derived from 
$${\cal B}_{\rm min}^{-1} -{\cal B}_{\rm max}^{-1}={2\beta_{r,\rm max}
\over\sqrt {1-\beta^2}}. \eqno (3)         
$$
It is easy to see that
$$\beta_{r,\rm max}=\beta\sin i.  \eqno (4)$$         
Therefore equations (2) and (4) yield
$$\sin i={{\cal B}_{\rm max}-{\cal B}_{\rm min}\over {\cal B}_{\rm max}
+{\cal B}_{\rm min}}~{1\over \beta},\eqno (5)$$                
where
$$\beta=\sqrt{1- \left({2{\cal B}_{\rm max}{\cal B}_{\rm min}\over 
{\cal B}_{\rm max}+{\cal B}_{\rm min}}\right)^2}. \eqno (6)$$                
Thus both the total velocity $\beta$ and the inclination angle $i$ are
derivable from the \pdm. 

In practical implementation of the above approach, we assume that the 
measured frequency $\tilde\nu$ is corrected for
annual motion of Earth (term $\cal E$). 
Motion of the visible star in the binary (with variable velocity $\vec\beta$) 
and motion of the binary barycentre (with constant velocity $\vec b$) 
result in a (variable) total Doppler shift 
$${\nu_0\over\tilde\nu}
= {1-\beta_r\over\sqrt {1-\beta^2}}{1-b_r\over\sqrt {1-b^2}} 
\approx  
\left(1-\beta_r \right)\left(1-b_r+{1\over 2}b^2+{1\over 2}\beta^2  
\right).(7)$$
The instants when ${\nu_0/\tilde\nu}$ reaches its maximum and minimum 
enable us to determine the instant when  $\beta_r=0$.
At that instant, observations bring us an approximate value
of $b_r$ (by neglecting the higher order terms ${1\over 2}b^2+
{1\over 2}\beta^2$).
Measurements of the binary's proper motion and distance yield the transverse 
velocity of the binary $b_t$ and thus enable one to evaluate $b^2=b_r^2+ 
b_t^2$.
The difference between $\nu_0/\tilde\nu$ (measured at the instant when
$\beta_r=0$) and $\left(1-b_r+{1\over 2}b^2
\right)$ yields an approximate value of ${1\over 2}\beta^2$. Obviously,
$\beta^2$ is kept constant in orbit. By applying the iteration method, the 
procedure is repeated until it converges to give accurate values of $b_r$. 
 As soon as $b_r,~ b^2$, and $\beta^2$ are found, the measured change of 
 $\nu_0/\tilde\nu$ in orbit yields $\vert\beta_r\vert$, the modulus of
$\beta_r$. Finally,
knowledge of both $\vert\beta_r\vert=\beta \sin i$ and $\beta^2$ allows us 
to evaluate $\sin i$. 
However, if $\beta \ll b$, which would correspond to wide binaries or a 
small mass of the companion, practical implementation of the above approach
is very difficult, as illustrated by case (ii).

\subsection{2.2. Case (ii): a single star with a planetary companion}  

Since the motion of a star that has a planetary companion orbiting around
it is only slightly disturbed by the planet, this case is reduced to case
(i) of a binary with $\beta\ll b$.
In the frame $S''$ associated with the barycenter of the Solar System 
$b$ is the speed of the source, $b_r$ is its radial component, and 
$b_t$ is the transverse component.
Although $\beta\ll b$, the disentangling of $\beta_r$ and $b_r$ does not make 
any problem as long as the terms linear in $\beta$ and  $b$ are only involved. 
However, in this case, as opposite to case (i), determinarion of the 
inclination angle of the planetary orbit turns out to be in practice very 
difficult. The reason is that inaccuracy in measuring $b_t$, the transverse
component of $b$, is currently too large. 
To see this, it is worth discussing the issue of accuracy  in more detail.

To account for the contribution of the second-order terms that 
enter the expansion of the relativistic Doppler formula into a series 
in $v/c$, precision of Doppler measurements of spectral line shifts 
(translated into the precision of radial velocity measurements) should be no 
less than
$$\eqalign{
\delta v &={1\over 2}v{v\over c}\cr 
&=67\hbox{ m/s if}~v=200\hbox{ km/s  (halo stars)}, \cr  
&=17\hbox{ m/s if}~v=100\hbox{ km/s  (bulge stars)}, \cr  
&=1.5\hbox{ m/s if}~v=30\hbox{ km/s  (disk stars).} ~~~~~~~~~~~~~~~~(8) \cr
}$$ 
Remarkably, accounting for the relativistic Doppler effect appears to be 
within the limits of modern astronomical techniques (see Sec. 3),
even when the measured velocity does not exceed 30 km/s.
            
In contrast to the radial velocity, the tangential velocity currently  cannot
be measured with a similar 
precision. However, since $v_t$ enters the relativistic Doppler equation
only as a second and higher order terms, accuracy in measuring proper motion 
and distance to the star are allowed to be rather modest. For instance,
if an inaccuracy in the inferred $v_t$ does not exceed, say, 10\%, 
as a product of combined uncertainties in measuring the distance to and 
proper motion of the star, a similar 
inaccuracy is translated into the value of total velocity, provided that 
the radial velocity measuremenets are as accurate as to obey equation (8). 
Although this would still make possible to evaluate the inclination angle 
of a binary as described by case (i), it fails for a planetary companion
because the current inaccuracy in 
measuring $b_t$ exceeds the value of $\beta$  
that has to be derived. For instance, even if $\delta V_t= 1~{\rm km~s^{-1}}$, 
the planetary companion induces $v$ typically not exceeding $0.1~{\rm 
km~s^{-1}}$. Nonetheless, since the ``signal" (the value of $v_t$) changes 
periodically, and the ``noise" (the value of $V_t$) is kept constant,
an extraction of that signal from the noise seems to be although a difficult 
but in principle solvable problem. 

\doublespace
\section{3. Precision Velocity Measurements}

As can be seen from equation (8),
the proposed method requires high precision in velocity measurements
as well as stability of the instrumental system. 
Any seasonal instrumental effects must be carefully eliminated. 
An idea of the appropriate reference frame 
for measuring velocities, namely passing the starlight 
through an absorbing gas prior to its entrance into the spectrograph, was
proposed long ago (Griffin \& Griffin 1973) but only recently has been 
developed to a level sufficient to provide high enough precision. The 
modern technique employs a stabilized gaseous iodine (I$_2$) absorption cell 
as a wavelength standard (Libbrecht 1988; Marcy \& Butler 1992; Cochran \& 
Hatzes 1994; K\"urster et al. 1994). The I$_2$ cell used in the wavelength 
region 5000-6000 ${\buildrel\circ\over {\rm A}}$, in combination with a high 
resolving power ${\cal R}$ provides a velocity precision proportional to 
${\cal R}^{-3/2}{\rm (S/N)}^{-1}$, where ${\rm S/N}$ is the signal-to-noise 
ratio (Hatzes \& Cochran 1994). The long-term velocity precision achievable 
with the current techniques is 4-7 m/s (K\"urster et al. 1994) to 3 m/s
or even better (Butler et al. 1996). For discussion of 
conditions that must be controlled in order to achieve such a high precision 
(changes in dispersion and the spectrograph 
point spread function, careful determination
of the topocentric velocity relative to the Solar system barycentre, etc.),   
see e.g. Marcy \& Butler (1992).
Other effects resulting in radial velocity changes, which have been 
revealed and studied during programmes of searches for extra-solar
planets, include the variability of K giants (\eg Hatzes \& Cochran 1994),
variability among non-Cepheid stars in the instability strip (Butler 1992),
and rapid oscillations of Ap stars (Hatzes \& K\"urster 1994).

Recently, the stabilized gas absorption cell technique has been implemented 
by several groups (Marcy \& Butler 1992; Cochran \& Hatzes 1994; K\"urster 
et al. 1994) in a search for  Jupiter-like planets. This procedure
is commonly named `precision radial velocity measurements', although, as is 
clear from equation (1), it is the changes in the {\it total} velocity 
(\ie including the transverse velocity) that are actually being measured
as soon as the accuracy of Doppler measurements reaches the level given by 
equation (8).

\section{4. Discussion}
%\doublespace
A central point of this letter is that if the accuracy of precision Doppler 
measurements is as high as to comply with equation (8), a combination (7) of
linear and quadratic in velocity terms, and not just radial velocity alone, 
is actually measured. If the radial velocity changes periodically (the source
is a binary star), the binary parameters, including the inclination angle  
of the orbit, can in principle be disentangled. In practice, this task
might be not easy. 
For instance, in an extreme case of a contact binary consisting of solar-mass
stars ($a\simeq 1.5~{\rm R_\odot}, ~P\simeq 0.2^{\rm d}$), the velocity 
vector changes in $\Delta t_{\rm sec}$  seconds by $\Delta v=(v^2/a)
\Delta t\simeq 0.14\Delta t_{\rm sec}$ m/s, i.e. by 1.4 m/s every 10 sec,
which requires a very short exposure time to make measurements.

Even if the source is a solitary star, precision Doppler measurements 
separated in time
can bring important dynamical information, especially if the star is located  
in a dense stellar field such as a globular cluster. Indeed, 
differentiation of equation (1) gives the frequency 
of a spectral line (corrected for annual motion of Earth) measured with 
time interval $\Delta t$ to be shifted by
$$ 
{\Delta\tilde\nu\over\tilde\nu}=
\left({\ddot {\vec V}\vec n\over c}+{V_t^2\over cd}+{V\ddot V\over c^2}
\right)\Delta t, \eqno (9)
$$
where $\ddot {\vec V}$ is acceleration and $d$ is distance to the source. 
A similar equation is used in pulsar 
timing (e.g. Blandford et al. 1993), where it is written in terms of pulsar 
period and its derivative.
In a globular cluster with the core radius $r_c=0.1$~pc and star density 
within the core $\rho_c= 3\cdot 10^6~{\rm M_\odot/pc^3}$, the dynamical
acceleration in the core is $\ddot V={4/3}\pi G\rho_c r_c=1.7\cdot 
10^{-5}~{\rm cm~s^{-2}}$. By neglecting the 2nd and 3rd terms in 
equation (9), one finds that in $\Delta t_{\rm yr}$ years the star's
velocity would change by $\Delta V=5.4 \Delta t_{\rm yr}$ m/s, i.e. 
by 5 m/s in a year of precision Doppler measurements. Therefore, they 
seem appropriate as a tool to measure gravitational field in the 
cores of globular clusters -- an opportunity that so far seemed only
possible for a few millisecond pusars. 

The prospects for application of the I$_2$ absorption cell technique (Sec. 3) 
seem to be very promising. Combining the echelle spectrograph with the 
I$_2$ cell allows for routine observations of stars of magnitude $V=13$ 
(Marcy \& Butler 1992). If the effects of all systematic errors 
can be reduced so that the velocity precision is limited only to the photon 
statistics, then the velocity precision improves as $N^{-1/2}$, $N$ being
the number of detected photons.
Fourier analysis makes it possible to extract the embedded periodicities 
with amplitudes of about 1 m/s (Brown 1990). This, 
as is seen from equation (8), enables one  
to measure the relativistic correction
to the ordinary Doppler effect even when peculiar velocities of stars
are $\sim 30$ km/s or less (disk stars).

Summarizing, this Letter  
demonstrates in principle a possibility to disentangle, for binary stars, 
the first and second order velocity terms of the relativistic Doppler effect
and, as a result, to extract the inclination angle of the orbit 
(by neglecting gravitational fields). This could be done by making use of 
recent substantial improvements in the technique of precision Doppler 
measurements combined with proper motion data.
However, much further work is needed to relax the 
underlying explicit and implicit assumptions 
and to evaluate the associated corrections as well.

\section*{Acknowledgments}
An earlier version of this Letter was written during my stay at 
Max-Planck-Institut f\"ur Extraterrestrische Physik in Garching-bei-M\"unchen. 
I am grateful to R. Genzel for hospitality and to  P. Schneider, N. Thatte, 
and S. White for helpful discussions.
My special thanks are due to S.M. Kopeikin, 
cooperation with whom has lead to a detailed 
relativistic theory of binary star orbits, with accounting for effects of 
general relativity (to be submitted).

\def\apj    {{ApJ{\rm,}\ }}

\def\aa     {{A\&A{\rm,}\ }}

\def\mnras  {{MNRAS{\rm,}\ }}
\def\ref#1  {\noindent \hangindent=24.0pt \hangafter=1 {#1} \par}
\def\v#1  {{{#1}{\rm,}\ }}

\section*{References}
\doublespace

\bye